\documentclass[%
 notitlepage,
reprint,
 amsmath,amssymb,
 aps,
prstab,
]{revtex4-2}

\usepackage{graphicx}
\usepackage{dcolumn}
\usepackage{bm}
\usepackage{hyperref}

\usepackage{booktabs} 
\usepackage{color}
\usepackage{multirow}
\usepackage{upgreek}

\newcommand{\KEK}{High Energy Accelerator Research Organization, Ibaraki 319-1106, Japan}

\begin{document}

\title{Quantitative Evaluation of a Hybrid Target for Ultraslow-Muon Production in $\mu$TRISTAN}

\author{Yasuhito Sakaki} \email{sakakiy@post.kek.jp}\affiliation{\KEK}
\author{Shusei Kamioka} \email{kamioka@post.kek.jp}\affiliation{\KEK}
\author{Mitsuhiro Yoshida} \affiliation{\KEK}

\date{\today}

\begin{abstract}
We present the first quantitative evaluation of the hybrid target concept proposed in $\mu$TRISTAN.
The $\mu$TRISTAN project is a positive-muon collider concept based on the ultraslow-muon production technique, in which thermal muonium emitted from a material surface is subsequently laser-ionized.
The hybrid target consists of a pion production target and a surrounding pion- and muon-stopping target that also serves as a muonium-production target.
In this paper, Monte Carlo simulations of the hybrid target are performed to evaluate the yield of positive muons stopped near the tungsten surface, where they can contribute to muonium emission. 
The ultraslow-muon yield at the hybrid target before extraction is estimated to reach the 10$^{-3}$ level per p--Li nuclear collision, assuming unit muon-to-muonium conversion efficiency.
This study provides a quantitative benchmark for hybrid-target design toward an intense ultraslow-muon source.

\end{abstract}

\maketitle


\section{Introduction}
Further progress in particle physics will likely require new collider facilities capable of extending precision studies of the Higgs sector and searches for physics beyond the Standard Model~\cite{EuropeanStrategyGroup:2020}. Among the various proposals for future collider facilities, a muon collider is considered one of the promising options~\cite{Accettura2023}. It offers the potential for both precision measurements and direct searches at the energy frontier. However, many of the key technologies required for a muon collider have not yet been established. One of the most important challenges is the production of a high-quality muon beam~\cite{Accettura2023}. 

Recently, $\mu$TRISTAN has been proposed~\cite{10.1093/ptep/ptac059}. 
The $\mu$TRISTAN project aims to realize a $\mu^+e^-$ or $\mu^+\mu^+$  collider to search for new physics as a Higgs factory or at the TeV scale by using only $\mu^+$. 
Various physics opportunities for the collider, including Higgs and supersymmetry studies, flavor physics, and dark-matter searches, have been explored~\cite{10.1093/ptep/ptac059,10.1007/JHEP062023086,10.1007/JHEP022024214,10.1007/JHEP022025116}.
One of the central ideas of this proposal is to extend the positive-muon cooling technique being developed at the Japan Proton Accelerator Research Complex (J-PARC), namely, the ultraslow-muon source~\cite{PhysRevLett.74.4811}. 
This source is being developed for the muon $g-2$/EDM experiment at J-PARC~\cite{10.1093/ptep/ptz030}, and the rf acceleration of ultraslow muons has recently been demonstrated~\cite{PhysRevLett.134.245001}.
To apply this ultraslow-muon technique to high-intensity applications requiring muon rates above $10^{10}$ muons/s, a new approach is needed to substantially increase the overall muon yield. 

As a possible solution to this problem, a new scheme has been proposed based on an energy-recovery internal-target ring, a hybrid target system consisting of a pion-production target surrounded by an ultraslow muon production target, and a second-stage cooling process~\cite{10.1093/ptep/ptac059}. 
The internal-target ring allows the circulating proton beam to interact repeatedly with the pion-production target, increasing the effective proton--nucleus collision rate. 
The produced pions are stopped in the surrounding target, and their daughter muons are used to generate thermal muonium, which is subsequently laser-ionized to produce ultraslow muons.
Since this source is spatially extended, the muons are extracted from the hybrid target and transported to the second-stage cooling system, where the ultraslow-muon production process is repeated to further compress the phase-space volume. After the second-stage cooling, the resulting muons are accelerated.
In this scheme, the hybrid target primarily increases the number of muons available for subsequent cooling, while the second-stage cooling further reduces the source size and phase-space volume before acceleration.

Among the elements of the proposed accelerator chain, the hybrid target plays a central role because its muon yield and source volume set the initial conditions for the beam performance that can ultimately be achieved.
So far, only a rough estimate of the ultraslow-muon yield has been made~\cite{10.1093/ptep/ptac059}. 
In addition, the characteristic size and spatial distribution of the muon-production region have not yet been quantitatively evaluated, although these quantities are important inputs for evaluating the extraction system and subsequent cooling stage.

In this paper, we present the first quantitative evaluation of the hybrid target system. The key figure of merit is the yield of muons stopped near the surface of the target material, since only these muons can contribute to the production of ultraslow muons. 
Using a Monte Carlo simulation based on a simplified geometry, we evaluate the yield of such surface-stopped muons per nuclear collision between an incident proton and the target material. 
We also derive a semi-analytical model describing muon diffusion and muonium emission.
We then combine the Monte Carlo results with this model to evaluate the target-geometry dependence of the ultraslow-muon yield at the hybrid target before extraction.

The remainder of this paper is organized as follows. Section~\ref{chap:usm} describes the basic properties of the ultraslow-muon source and its current limitations. Section~\ref{chap:target-design} introduces the concept of a new ultraslow-muon production target. Section~\ref{chap:phits-sim} presents a Monte Carlo evaluation of the ultraslow-muon yield before extraction. Section~\ref{chap:future} discusses future prospects, followed by the conclusions.

\section{Current ultraslow-muon source and its limitations}
\label{chap:usm}

Ultraslow muons are generated via resonance-enhanced multiphoton laser ionization of thermal muonium, a bound state of a positive muon and an electron, generated from a muonium production target. 
Being a thermal particle source, the initial transverse phase-space is determined primarily by the thermal velocity spread and the source size. A transverse rms emittance of a thermal particle source can be estimated as follows~\cite{Reiser2008}:
\begin{equation}
\label{eq:thermal_emittance}
    \epsilon_{n,\rm{rms}} \sim \sigma_s\sqrt{\frac{kT}{mc^2}},
\end{equation}
where $m$ is the mass of the particle, $T$ is the temperature, and $\sigma_s$ is the transverse rms size of the source. Assuming a typical rms source size of 5~cm, $\epsilon_{n,\rm{rms}}$ is estimated to be $ \sim 1~\pi$mm$\cdot$mrad for a room-temperature target.

In the present scheme, a surface-muon beam is used as the incident beam to a muonium production target because it provides a low-energy ($E_{\rm{kin}}\sim 4$~MeV), quasi-monoenergetic ($\Delta p/p\sim 5\%$), and nearly fully polarized source of positive muons.
Surface muons are produced by the decay at rest of $\pi^+$ stopped near the surface of the pion-production target.
At J-PARC, the yield of $\pi^+$ stopping within 0.2~mm of the target surface is estimated to be $10^{-5}$ per incident proton~\cite{10.1093/ptep/pty116}. 

Several muonium-emitter materials have been demonstrated, including hot tungsten~\cite{PhysRevLett.56.1463} and silica aerogel~\cite{10.1093/ptep/ptu116,10.1093/ptep/ptaa145}, which emit thermal muonium into vacuum from stopped muons in the material. 
For an aerogel target, the muonium emission efficiency relative to the incident surface-muon flux is estimated to be 1.3\%, including a muonium formation efficiency of 52\%~\cite{ZHANG2022167443}. The overall efficiency is limited because only muons stopped within a shallow region near the surface can escape into the vacuum before they decay while the stopping distribution of the incoming muons is widely spread inside the target~\cite{ZHANG2022167443}. 
At the time of laser irradiation, about 30\% of the produced muonium atoms can overlap with the laser field.
The ionization efficiency for a single muonium can be as high as 73\%~\cite{10.1093/ptep/ptz030}.

The present scheme is subject to two successive surface-limited processes: the surface-muon yield is limited by the number of pions stopping near the production-target surface, and the ultraslow-muon yield is further limited by the number of muons stopping near the emitter surface. Therefore, simply increasing the accepted muon flux, for example by broadening the momentum acceptance of the transport line, does not directly lead to a proportional increase in the ultraslow-muon intensity.

A multilayer muonium-emitter configuration has been proposed to increase the effective emitting surface area~\cite{ZHANG2022167443,Bai:2024skk}. While this scheme is expected to provide a multiplicative improvement in the conversion efficiency, it does not address the fundamental limitation imposed by the low surface-muon production yield.

In contrast, once a sufficiently intense first-stage ultraslow-muon source is available, a subsequent ultraslow-muon production stage can be used as an additional cooling step, provided that the incident muon beam has a sufficiently narrow momentum spread and can be focused onto a small spot at the muonium-production target.
In this case, the stopping distribution can be localized near the emitter surface, improving the fraction of muons that contribute to muonium emission, while the transverse emittance of the re-emitted thermal source is reduced according to Eq.~\ref{eq:thermal_emittance} through the smaller source size. 
In an idealized limit, the conversion efficiency of such a stage is expected to be limited by the muonium-formation probability and by the fraction emitted into the vacuum side, giving an order-of-magnitude efficiency of $\sim 25\%$ before laser-ionization and extraction losses.
Thus, the first-stage source must provide a sufficiently large muon rate.

\begin{figure*}[t]
  \centering
  \includegraphics[width=0.8\linewidth]{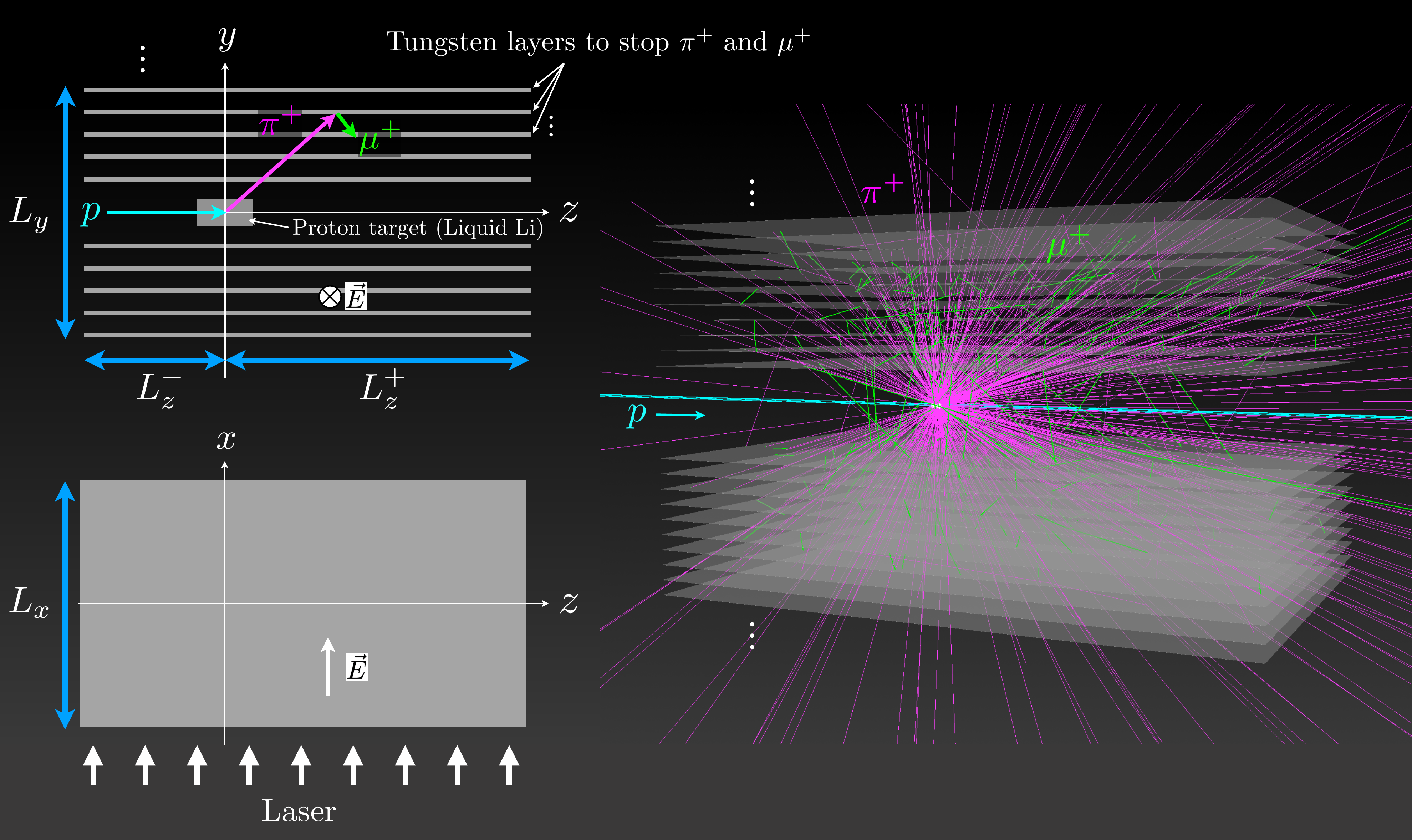}
  \caption{
    Conceptual view of the hybrid target.
    The target consists of a liquid-lithium target surrounded by multilayer tungsten structures. 
    Each tungsten layer consists of an array of parallel tungsten wires lying in the same plane. 
    The beam direction is taken to be the $z$-axis.
    Pions are produced at the lithium target. These pions are stopped inside the tungsten layers, followed by the emission of decay muons. The muons are also stopped inside the tungsten and thermal muonium atoms are emitted into vacuum.
    Laser pulses are introduced into each interlayer gap along the $x$-axis. The resulting ultraslow muons are extracted along the $x$-axis.
    }
  \label{fig:concept}
\end{figure*}

\section{Hybrid target concept and modeling assumptions}
\label{chap:target-design}
\subsection{Overview}
\label{sec:concept}

The proposed hybrid target consists of a liquid-lithium target surrounded by multilayer tungsten structures, as illustrated in Fig.~\ref{fig:concept}. In this paper, we adopt $E_{\rm beam}=3$~GeV as the baseline proton-beam energy, following the $\mu$TRISTAN proposal~\cite{10.1093/ptep/ptac059}. The role of the lithium target is to produce $\pi^+$, while the surrounding tungsten layers stop the produced $\pi^+$ and the daughter $\mu^+$, so that the stopped $\mu^+$ can eventually be emitted as muonium into vacuum. We refer to this combined system as the hybrid target.
Each tungsten layer consists of an array of parallel tungsten wires lying in the same plane. An electric field for drifting the resulting ultraslow muons is generated by applying different voltages to the neighboring wires. 
A fraction of the proton beam undergoes nuclear collisions in the target, while the remaining protons are recirculated in the ring and directed back to the target.

The coordinate system is defined as follows.
The beam direction is taken to be the $z$-axis, the stacking direction of the tungsten layers the $y$-axis, and the muon extraction-field direction the $x$-axis. The liquid-lithium target is located at the origin.
We introduce four parameters to characterize the geometrical size of the hybrid target system: $L_x$ and $L_y$ denote the total lengths along the $x$-and $y$-axes, respectively, while $L_z^-$ and $L_z^+$ denote the upstream and downstream extents of the target measured from the origin along the $z$-axis.
Along the $x$ and $y$-directions, the target is taken to be symmetric about the origin, with $-L_x/2\le x \le L_x/2$ and $-L_y/2 \le y \le L_y/2.$
An asymmetric layout of the tungsten stopping region is assumed along the z direction, with ~$L_z^{+} > L_z^{-}$. The $\pi^+$ produced by the proton beam are emitted over a broad angular range. The higher-energy component, however, is more strongly forward-directed and has a longer range. Therefore, a longer stopping region is required on the downstream side than on the upstream side. As a result, extending the stopping region preferentially on the downstream side is an efficient way to keep the target system compact.

\subsection{Pion production target}

In this work, liquid lithium, a low-Z material, is adopted as the pion-production target. Its advantages for use as an internal target in an energy recovery ring are its ability to withstand a megawatt-class proton beam, the small emittance growth caused by multiple Coulomb scattering, and the small energy loss per pass, which is favorable for efficient beam recovery. The suitability of liquid lithium for such applications has been demonstrated in previous studies, including those by the IFMIF and CiADS collaborations~\cite{Knaster2017,Cai2024}.

The optimal liquid-lithium target thickness depends on the recirculating proton-ring configuration and beam-recovery requirements. We therefore evaluate the surface-stopped muon yield per p--Li nuclear collision as the key figure of merit for the hybrid target system, thereby isolating its target-side performance.

\subsection{Pion stopping, muon thermalization, and muonium emission}
\label{sec:stopping-emission}
The $\pi^+$ emitted from the liquid-lithium target are decelerated in the surrounding multilayer tungsten structure, and a fraction of them stop there. The stopped $\pi^+$ decay at rest and emit low-energy $\mu^+$ isotropically, with an energy comparable to that of conventional surface muons.
The emitted $\mu^+$ are also decelerated in the tungsten and stop there. The stopped $\mu^+$ then diffuse thermally inside the tungsten, and those that reach the tungsten surface before decaying are emitted as thermal muonium. 
The pion lifetime of 26~ns is an order of magnitude shorter than the other relevant timescales, including the muon lifetime and the muonium-emission timescale. In the present study, therefore, all stopped $\mu^+$ are assumed to be generated simultaneously.
Although thermal muonium emission from hot tungsten has been observed, direct emission of bare thermal $\mu^+$ has not been reported. 
The probability that a surface-reaching $\mu^+$ is emitted as muonium has not, however, been separately quantified.
We therefore treat this probability as a material-dependent conversion factor, $\eta_{\rm conv}$, in the yield estimate below.

As discussed in Appendix~\ref{app:diff}, we define the characteristic diffusion length as
\begin{equation}
    \ell_{\rm diff} = \sqrt{D \tau_{\mu}},
\end{equation}
where $D$ is the effective diffusion coefficient of stopped $\mu^+$ in tungsten and $\tau_{\mu}$ is the muon lifetime. This length represents the characteristic depth from which stopped $\mu^+$ can reach the surface before decaying. 
The probability that a stopped $\mu^+$ initially located at a depth $z_0$ reaches the surface before decaying is given by
\begin{equation}
    P(z_0) = \exp(-z_0/\ell_{\rm diff}).
\end{equation}
For the numerical evaluation, we assume a hot-tungsten temperature of $T=2000$~K.
The diffusion coefficient is reported to be $2\times10^{-3}\,\mathrm{cm^2/s}$~\cite{PhysRevLett.56.1463}. 
Using this value together with the muon lifetime, we obtain $\ell_{\rm diff}\approx 0.7~\upmu\mathrm{m}$. 
For an approximately uniform stopping distribution near the surface, 
approximately 95\% of the emitted muonium atoms originate from muons stopped 
within $3\,\ell_{\rm diff}$ of the surface.

\subsection{Drift field and the required interlayer gap}
\label{sec:drift-gap}
Laser pulses are introduced into each interlayer gap along the $x$-axis to ionize the muonium in the gap, as assumed in the previously proposed multilayer target system~\cite{ZHANG2022167443}.
Muonium is gradually emitted into vacuum, and the emission timescale is determined by the muon lifetime and the diffusion coefficient. 
After emission, the muonium travels through the space between neighboring layers. 
In the present study, a muonium atom is assumed to be lost once it collides with a neighboring layer. 
To the best of our knowledge, no quantitative model for muonium collisions with tungsten surfaces under the present conditions is available. This assumption is therefore adopted as a conservative baseline.

\begin{figure}[t]
  \centering
  \includegraphics[width=0.9\linewidth]{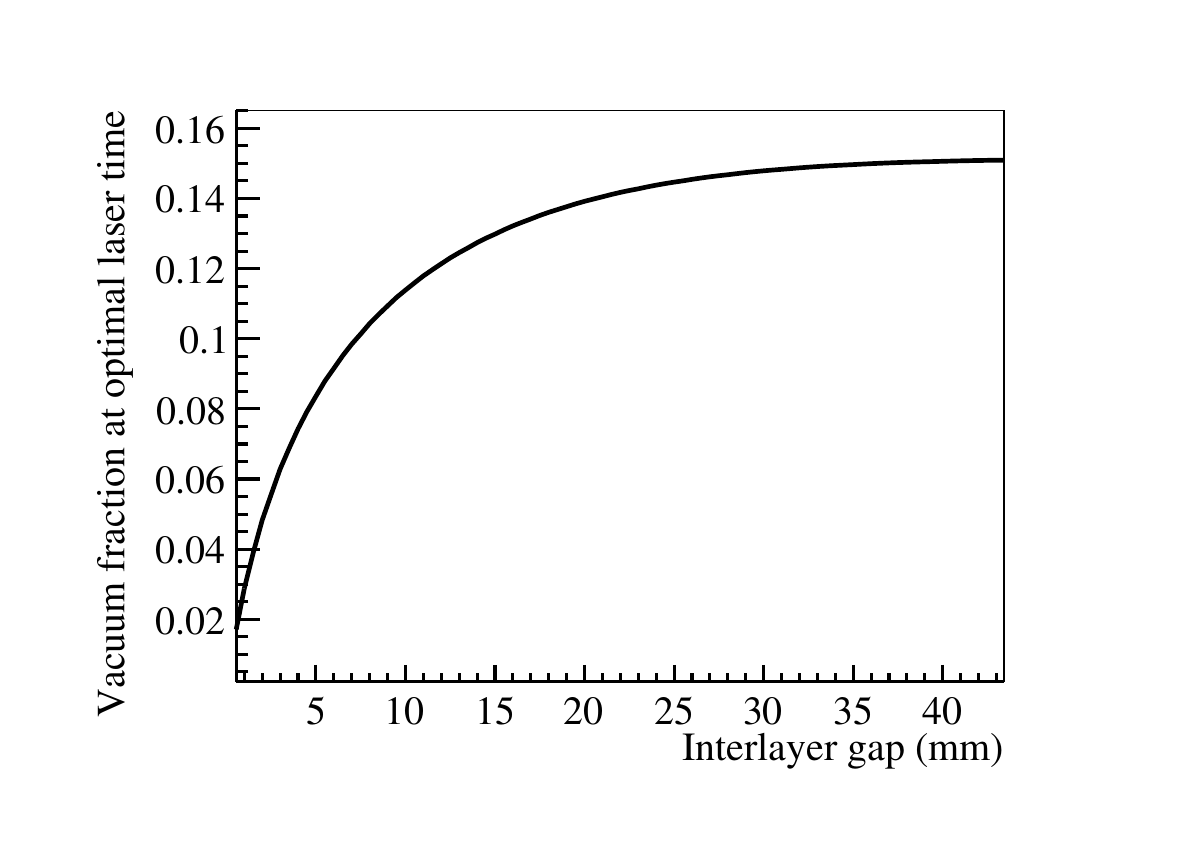}
  \caption{
  Fraction of muonium present in vacuum at the optimal laser-irradiation time as a function of the interlayer gap.
  The calculation assumes hot tungsten with $D=2\times10^{-3}\,\mathrm{cm^2/s}$ and $T=2000~\mathrm{K}$.
  A uniform stopping distribution near the surface and unit muon-to-muonium conversion efficiency at the surface are assumed. The vertical axis is normalized to the number of muons stopped within $3\ell_{\rm diff}$ of the surface.
  }
  \label{fig:emission-fraction}
\end{figure}

For each interlayer gap, the optimal laser-irradiation time, at which the largest fraction of muonium atoms exists between neighboring layers, and the corresponding vacuum fraction can be estimated from the model discussed in Appendix~\ref{app:diff}.
We denote this gap-dependent fraction by $\eta_{\rm vac}^{\rm laser}$.
It is normalized to the number of muons stopped within $3\,\ell_{\rm diff}$ from the surface, assuming unit muon-to-muonium conversion efficiency and uniform stopping distribution near the surface.
Figure~\ref{fig:emission-fraction} shows $\eta_{\rm vac}^{\rm laser}$ as a function of the interlayer gap.
Approximately 10\% of the muons stopped within $3\,\ell_{\rm diff}$ of the surface can be present in vacuum as muonium at the optimal laser-irradiation time.
A larger gap increases the fraction of muonium remaining in vacuum, but also requires a larger target system because fewer tungsten layers can be accommodated per unit length.
We therefore evaluate the ultraslow-muon yield for four representative gaps, 5, 10, 15, and 20~mm, to compare the performance of the hybrid target system.

The resulting ultraslow muons are driven along the $x$ direction and extracted from the target system by applying a voltage difference across the tungsten wires. A fraction of the muons is lost during the drift because of muon decay and collisions with the tungsten layers. The detailed design of the drift system and a quantitative evaluation of the associated muon loss are beyond the scope of this work and are not discussed here.

\section{Simulation studies of the proposed target}
\label{chap:phits-sim}
\subsection{Simulation framework}

In the present work, we employ the Particle and Heavy Ion Transport code System (PHITS) for Monte Carlo simulations~\cite{Sato:2023qsv}.
We adopt the JAM model~\cite{Nara2000} as the hadronic-physics model and apply an angle-dependent correction to the $\pi^+$ production cross section based on a comparison with the HARP data~\cite{HARP2008}.
Details of the model choice and the correction are given in Appendix~\ref{app:model-comparison}.

\subsection{Geometrical modeling}
For the range of target thicknesses considered in this study, the fractional energy loss of the proton beam within the target is calculated to be less than $\sim 10^{-3}$. Therefore, the pion-production cross section and the pion kinematics are essentially independent of the target thickness and size.
Accordingly, the liquid-lithium target is approximated as a point-like pion-production source placed at the origin.
We characterize the pion and muon yields per p--Li nuclear collision in the target. Here, following the collision-length convention, a p--Li nuclear collision includes both elastic and inelastic scatterings, because either type of scattering is assumed to remove the circulating proton from the stored beam. 
With this normalization, the result does not depend on the target thickness.
In addition, although each tungsten layer is conceptually composed of an array of tungsten wires, it is approximated as a single plane with a dimension~$L_{x}\times(L_{z}^++L_{z}^-)$.
Muons that have kinetic energies below $1~\mathrm{keV}$ inside the target are identified as stopped muons.  

\begin{figure*}[t]
  \centering
  \includegraphics[width=0.7\linewidth]{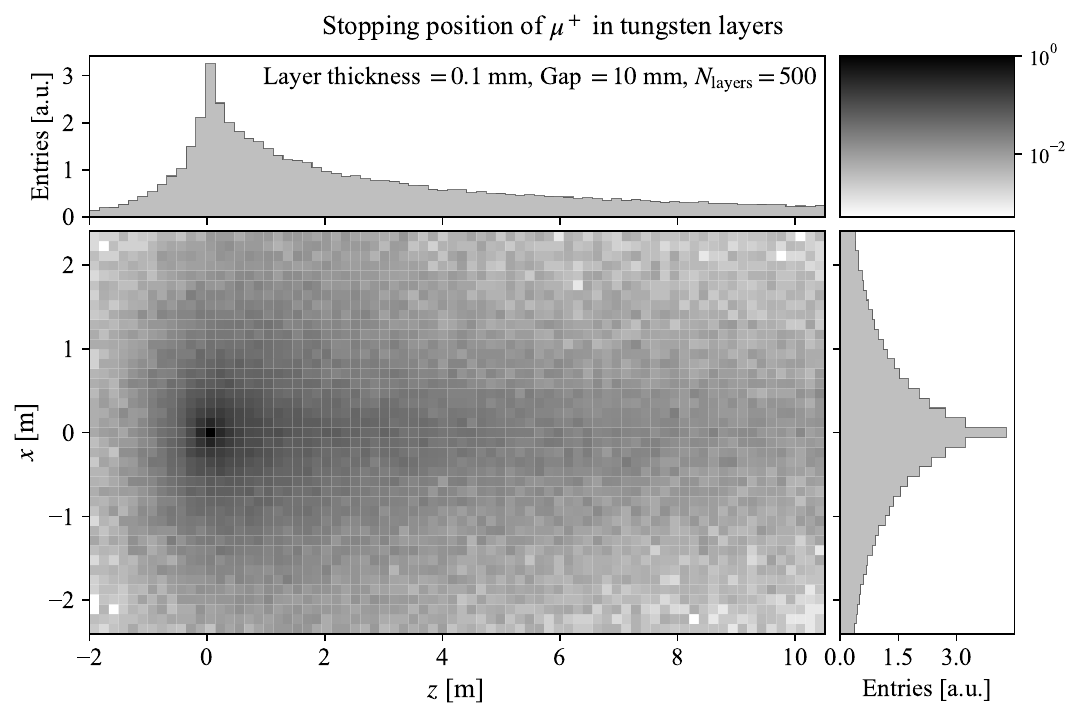}
  \caption{
    Stopped-muon distribution inside the tungsten layers.
    Central panel: two-dimensional distribution on a logarithmic scale. Upper and right panels: projections onto the $z$ and $x$ axes, respectively.
    The layer thickness is $0.1$~mm, the interlayer gap is $10$~mm, and the number of layers is $N_{\rm layers}=500$.
    }
  \label{fig:muon-xz}
\end{figure*}

Performing a full Monte Carlo simulation for every possible target geometry
and the tungsten-layer parameter set would be computationally expensive.
Therefore, we employ the following simplified treatment.

First, we perform a simulation with a large homogeneous tungsten geometry in which the tungsten density is reduced from its physical value \(\rho_{\rm W}\) to a reference density \(\rho_{\rm W,mod}\), and record the muon stopping positions.
In this work we set \(\rho_{\rm W,mod}=\rho_{\rm W}/50\).
This reduced density is used as a numerical reference of the same order as the volume-averaged tungsten densities of the representative multilayer geometries considered below, so that the subsequent density rescaling remains modest.
For a multilayer geometry with tungsten-layer thickness \(d_{\rm W}\) and interlayer gap \(d_{\rm gap}\), the corresponding volume-averaged tungsten density is
\[
\rho_{\rm eff}
=
\rho_{\rm W}\frac{d_{\rm W}}{d_{\rm W}+d_{\rm gap}} .
\]
We map the stopping positions obtained in the reduced-density bulk simulation to each multilayer geometry by preserving the tungsten areal density.
The density-scaling factor relative to the simulated bulk density is
\[
s_\rho
=
\frac{\rho_{\rm eff}}{\rho_{\rm W,mod}}
=
\frac{d_{\rm W}}{d_{\rm W}+d_{\rm gap}}
\frac{\rho_{\rm W}}{\rho_{\rm W,mod}} .
\]
Thus, transport lengths in the reduced-density bulk simulation are multiplied by \(1/s_\rho\) when they are mapped to the corresponding multilayer geometry.

This treatment should be regarded as an effective-density, coarse-grained approximation.
It assumes that the vacuum gaps contribute negligibly to the material depth and that their main effect is to reduce the average tungsten areal density along particle trajectories.
The stopping phase within the periodic multilayer structure is treated as effectively averaged after integrating over pion-production kinematics, emission angles, and range straggling.
It also assumes that the muonium-emission probability is determined mainly by the distance from the stopping point to the nearest tungsten surface.
We validated this simplified treatment by comparing the total stopped-muon yield obtained with the reduced-density bulk approach to that of a full multilayer-geometry simulation for a representative parameter set with a layer thickness of $0.1$~mm, an interlayer gap of $10$~mm, $N_{\rm layers}=200$, $L_x=3$~m, $L_z^{-}=1$~m, and $L_z^{+}=10$~m. The two results agreed within the statistical uncertainty of the simulation, which is at the level of a few percent.

\subsection{Results and target-geometry dependence}

Figure~\ref{fig:muon-xz} shows the $x$--$z$ distribution of stopped muons in the tungsten layers.
The layer thickness is $0.1$~mm, the interlayer gap is $10$~mm, and the number of layers is $N_{\rm layers}=500$.
The central panel shows the two-dimensional distribution on a logarithmic scale, while the upper and right panels show the projections onto the $z$ and $x$ axes, respectively.
The stopped-muon distribution has a sharp peak near the origin along the $z$ direction and exhibits an asymmetric shape with a long tail extending toward $z>0$. 
This reflects the fact that higher-momentum $\pi^+$ with longer ranges are preferentially emitted in the forward direction.
The distribution extends to $z\sim 10$~m, supporting the asymmetric target design discussed in Sec.~\ref{chap:target-design}. 
In contrast, the distribution along the $x$ direction is nearly symmetric about the origin.
To characterize the intrinsic spatial spread of the stopped-muon distribution, before imposing the reference target boundaries, the one-sided rms extents for \(z>0\) and \(z<0\) are evaluated to be 10.9~m and 1.3~m, respectively, while the rms extent along the \(x\) direction is 2.7~m.
The distribution along the $y$ direction (the stacking direction) is intrinsically discontinuous because muons can only stop within the tungsten layers. When coarse-grained over the layer-gap period, however, it is nearly identical to the $x$ distribution, reflecting the approximate azimuthal symmetry of the pion emission about the beam axis.
In this work, we adopt $L_z^+=10$ m, $L_z^-=1$ m, and $L_x=3$ m as a representative target geometry reflecting the characteristic spatial extent and forward asymmetry of the stopped-muon distribution.
This distribution, together with the total extent of the multilayer structure along the stacking direction, provides an estimate of the initial spatial extent of the ultraslow muons produced in the hybrid target. 
Combined with the thermal velocity distribution and the angular distribution of emitted muonium, the spatial distribution obtained here provides the basis for estimating the initial transverse emittance once the extraction geometry is specified.

Hereafter, we assume that, on the micrometer scale relevant to muonium emission, the stopped-muon distribution is approximately uniform within each tungsten layer near the surface, since the stopping phase relative to the multilayer period is expected to be effectively averaged.

The ultraslow-muon yield at the hybrid target before extraction can be calculated as
\begin{equation}
Y_{\mathrm{USM}}^{\mathrm{total}}
=\eta_{\mathrm{REMPI}}\times \eta_{\mathrm{conv}}\times \eta_{\mathrm{vac}}^{\mathrm{laser}}\times \eta_\mu,
\end{equation}
Here, $\eta_{\mu}$ denotes the yield of positive muons stopped within 
$3\,\ell_{\rm diff}$ of a tungsten surface per p--Li nuclear collision. 
The factor $\eta_{\rm vac}^{\rm laser}$ is the fraction of muonium present
in the interlayer vacuum at the optimal laser-irradiation time, as defined
in Sec.~\ref{sec:drift-gap}. The factor $\eta_{\rm conv}$ is the material-dependent
muon-to-muonium conversion factor introduced in Sec.~\ref{sec:stopping-emission}. 
Finally, $\eta_{\mathrm{REMPI}}=0.73$ is the laser-ionization efficiency for 
thermal muonium~\cite{10.1093/ptep/ptz030}.

Using the Monte Carlo simulation together with the semi-analytical emission model described in Appendix~\ref{app:diff},
we evaluate the product $\eta_{\mathrm{vac}}^{\mathrm{laser}}\eta_\mu$.
For each set of tungsten-layer thickness, interlayer gap, and number of layers $N_{\mathrm{layers}}$,
the stopped-muon yield $\eta_\mu$ is obtained from the Monte Carlo simulation,
whereas $\eta_{\mathrm{vac}}^{\mathrm{laser}}$ is determined by the interlayer gap through the optimal laser-irradiation time.

\begin{figure*}[t]
  \centering
  \begin{minipage}[t]{0.495\linewidth}
    \centering
    \includegraphics[width=\linewidth]{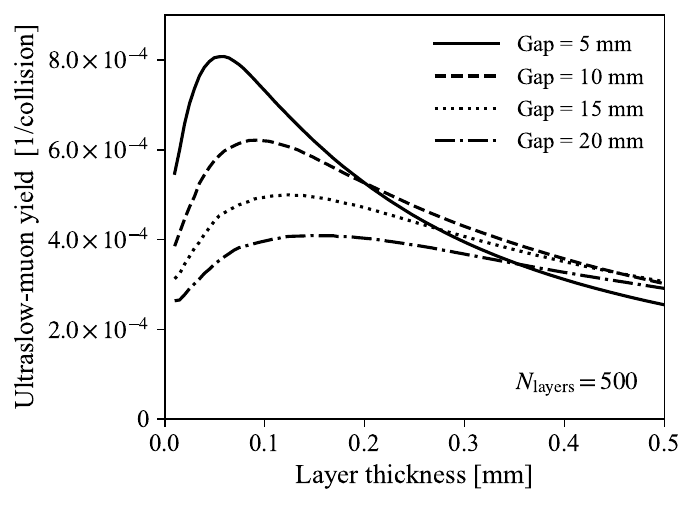}\\
    (a)
  \end{minipage}%
  \hspace{0.005\linewidth}%
  \begin{minipage}[t]{0.495\linewidth}
    \centering
    \includegraphics[width=\linewidth]{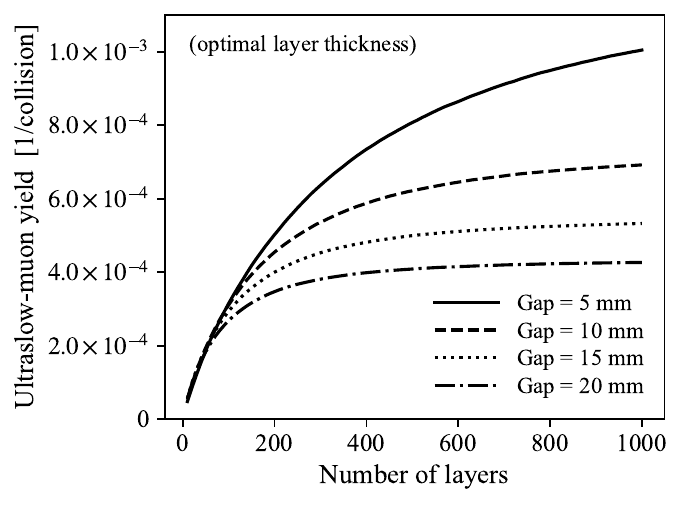}\\
    (b)
  \end{minipage}
  \caption{
    Ultraslow-muon yield before extraction per p--Li nuclear collision at the hybrid target for different interlayer gaps, assuming unit muon-to-muonium conversion efficiency. The proton beam energy is 3~GeV.
    (a) Yield as a function of layer thickness for $N_{\mathrm{layers}}=500$.
    (b) Yield as a function of $N_{\mathrm{layers}}$, with the layer thickness fixed to the optimal value for each gap.
    }
  \label{fig:Y-gap}
\end{figure*}

Figure~\ref{fig:Y-gap} shows the results of the parameter scan.
The vertical axis represents $\eta_{\mathrm{REMPI}}\eta_{\mathrm{vac}}^{\mathrm{laser}}\eta_\mu$, namely the ultraslow-muon yield per p--Li nuclear collision for unit muon-to-muonium conversion efficiency.
Figure~\ref{fig:Y-gap}(a) shows the yield as a function of the tungsten-layer thickness for four representative interlayer gaps. The number of layers is fixed to 500.
This result shows that there is an optimal layer thickness for the given gap.
Increasing the layer thickness improves the stopping efficiency for pions and daughter muons, but decreases the fraction of stopped muons located near the tungsten surface.
On the other hand, if the layer thickness is too small, the total tungsten thickness becomes insufficient for stopping pions and daughter muons when the number of layers is fixed.
Figure~\ref{fig:Y-gap}(b) shows the yield as a function of the number of layers for four representative gaps, with the layer thickness fixed to the optimal value for each gap.
Increasing the number of layers improves the yield, and the yield saturates at approximately 0.001 when the gap is 5~mm once the total target thickness becomes sufficiently large to stop the relevant pions and daughter muons.

The realistic optimum values of these parameters and the overall geometry should be determined together with the design of the ultraslow-muon extraction system, the laser system, and the lithium target. 
Nevertheless, the present results provide the first quantitative benchmark for the target-side performance of the hybrid target.

\subsection{Estimate of pre-extraction ultraslow-muon yield}
Finally, we provide an order-of-magnitude estimate of the ultraslow-muon intensity before extraction, illustrating how the target-side yield may translate into the actual muon intensity.
We consider the idealized upper-limit case in which each injected proton is eventually lost through one p--Li nuclear collision.
In that case, the number of ultraslow muons produced at the hybrid target before extraction can be estimated as
\begin{equation}
N_{\mathrm{tot}}
=\eta_{\mathrm{REMPI}}\times \eta_{\mathrm{conv}}\times \eta_{\mathrm{vac}}^{\mathrm{laser}}\times \eta_\mu \times N_p,
\end{equation}
where $N_p$ is the number of protons per second. 
Table~\ref{table:ref_value} summarizes the target geometry used for this estimation.
For a 1 MW, 3 GeV proton beam, $N_p = 2\times 10^{15}\ \mathrm{s}^{-1}$. This gives
\begin{equation}
N_{\mathrm{tot}}
=\eta_{\mathrm{conv}}\times 1.2\times 10^{12}\times \left(\frac{\mathrm{Beam\ Power}}{1\ \mathrm{MW}}\right)\ \mathrm{s}^{-1}.
\end{equation}  

This estimate represents the ultraslow-muon production rate available at the hybrid target before extraction under the idealized assumption described above.
Determining the extracted muon rate will require a coupled evaluation of the proton-ring design, beam-recovery efficiency, and extraction efficiency.

\begin{table}[b]
\caption{
Representative high-yield parameter set used for the order-of-magnitude
pre-extraction intensity estimate. The muon-to-muonium conversion efficiency
$\eta_{\rm conv}$ is not included in the quoted ultraslow-muon yield.
The yield carries a systematic uncertainty of $\sim 10\%$ from the hadronic-model treatment (see Appendix~\ref{app:model-comparison}).
}
\begin{ruledtabular}
\begin{tabular}{lc}
Quantity & Value \\
\hline
$L_x$ & $3~\mathrm{m}$ \\
$L_z^+$ & $10~\mathrm{m}$ \\
$L_z^-$ & $1~\mathrm{m}$ \\
Interlayer gap & $10~\mathrm{mm}$ \\
Layer thickness & $0.1~\mathrm{mm}$ \\
Number of layers, $N_{\rm layers}$ & 500 \\
$L_y$ & $\sim 5~\mathrm{m}$ \\
Ionization efficiency, $\eta_{\rm REMPI}$ & 0.73 \\ \hline
Ultraslow-muon yield, $\eta_{\rm REMPI}\,\eta_{\rm vac}^{\rm laser}\,\eta_\mu$
& $6\times10^{-4}$ \\
\end{tabular}
\end{ruledtabular}
\label{table:ref_value}
\end{table}

\section{Future prospects}
\label{chap:future}
This study has focused on the conceptual design of the hybrid target and on the evaluation of the muon stopping efficiency near the target surface. 
The present benchmark also quantifies the large spatial scale of the ultraslow-muon production region, which provides an important constraint for the design of the extraction and subsequent cooling systems.
As the next step, the design and simulation of ultraslow-muon extraction and the second cooling stage should be pursued.
Such a study will also be needed to determine the optimal target dimensions.
A subsequent step will be the design and simulation of the recirculating proton ring and the optimization of the lithium-target thickness, including beam-recovery considerations. 
These studies will provide the basis for a future
end-to-end estimate of the ultraslow-muon intensity and emittance. 
Engineering issues related to practical implementation, such as the radiation environment associated with a megawatt-class proton beam, the detailed design of the tungsten layers, and the design of the lithium target, should also be addressed.

\section{Conclusions}
\label{chap:conclusion}

In this work, we presented a quantitative evaluation of the hybrid target concept for $\mu$TRISTAN based on Monte Carlo simulations.
The key figure of merit for ultraslow-muon production is the yield of muons stopped near the surface of the muonium-production target, since only these muons can contribute to muonium emission into vacuum.
Using a simulation framework based on the JAM hadron model with a correction to the pion-production kinematics derived from comparisons with HARP data, we evaluated the stopped-muon yield for the proposed target geometry.
As a result, assuming unit muon-to-muonium conversion efficiency, the ultraslow-muon yield at the hybrid target before extraction was found to reach the 10$^{-3}$ level per p--Li nuclear collision.
The characteristic spatial extent of the hybrid-target source was also quantified, revealing a meter-scale, forward-asymmetric source region. 
Together,  these results provide a quantitative benchmark for the hybrid-target design in $\mu$TRISTAN.

\begin{acknowledgments}
This work was supported in part by JSPS Kakenhi Grant Numbers 24K00650 and 23K13131.

\end{acknowledgments}

\appendix

\section{Diffusion model for muonium emission and spatial distribution}
\label{app:diff}
Here we present a semi-analytical model for the diffusion of muons inside the target and their motion after emission, instead of a full simulation treatment. 

We consider a muon stopped inside a material at a depth $x_0>0$ from the surface, where the surface is located at $x=0$.  
The diffusion toward the surface is approximated as one-dimensional Brownian motion with an effective diffusion coefficient $D$ for stopped $\mu^+$ in tungsten. 
In the present study, all stopped $\mu^+$ are assumed to be generated simultaneously at $t=0$ as the pion lifetime of 26 ns is an order of magnitude shorter than the other relevant timescales. As in the main text, we assume that the stopped-muon distribution is approximately uniform over the depth scale relevant to muonium
emission.

For a particle initially located at $x_0$, the first-passage-time density to the surface is given by
\begin{equation}
f(t\mid x_0)
=
\frac{x_0}{\sqrt{4\pi D\,t^3}}
\exp\!\left(-\frac{x_0^2}{4Dt}\right),
\qquad t>0.
\end{equation}

When the muon lifetime is taken into account, the probability that a stopped $\mu^+$ initially located at depth $x_0$ reaches the surface before decay is
\begin{equation}
P_{\rm emit}(x_0)
=
\int_0^\infty e^{-t/\tau_\mu} f(t\mid x_0)\,dt
=
\exp\!\left(-\frac{x_0}{\sqrt{D\tau_\mu}}\right),
\end{equation}
where $\tau_\mu$ is the muon lifetime.  
Thus, the characteristic attenuation length for emission is $\sqrt{D\tau_\mu}$. Approximately 95\% of the emitted particles originate from within a depth of $3\sqrt{D\tau_{\mu}}$ from the target surface. 
When the initial particles are uniformly distributed within $3\sqrt{D\tau_{\mu}}$ from the surface, approximately one third of them are emitted to the vacuum before decay.

If the initial depth distribution is given by $\rho(x_0)$, the surface-emission-time density before including decay is
\begin{equation}
F_0(t)
=
\int_0^\infty \rho(x_0)\, f(t\mid x_0)\,dx_0,
\end{equation}
and the corresponding density including decay is
\begin{equation}
F_{\rm emit}(t)
=
e^{-t/\tau_\mu}F_0(t).
\end{equation}

If the initial distribution is uniform in the interval $0\le x_0\le L$,
\begin{equation}
\rho(x_0)=
\begin{cases}
1/L & (0\le x_0\le L),\\
0   & \text{otherwise},
\end{cases}
\end{equation}
then
\begin{equation}
F_0(t)
=
\frac{1}{L}\sqrt{\frac{D}{\pi t}}
\left(1-e^{-L^2/(4Dt)}\right).
\end{equation}

After reaching the surface, the emitted muonium is assumed to move ballistically in vacuum.  
Hereafter, we assume that the surface of the target lies at $x=0$ and the muonium target exists at $x<0$. 
We consider the spatial distribution along the surface normal direction $x>0$. Muonium is emitted towards $x>0$.
If the normal velocity component $v_x$ obeys a flux-weighted thermal distribution,
\begin{equation}
g(v_x)
=
\frac{m}{kT}v_x
\exp\!\left(-\frac{m v_x^2}{2kT}\right),
\qquad v_x>0,
\end{equation}
where $m$ is the muonium mass and $T$ is the temperature, then the density at position $x$ and time $t$ is given by
\begin{equation}
n(x,t)
=
e^{-t/\tau_\mu}
\int_0^t dt_c\,
\frac{1}{t-t_c}\,
F_0(t_c)\,
g\!\left(\frac{x}{t-t_c}\right).
\end{equation}
Here, $t_c$ denotes the time at which the particle reaches the surface and is emitted into vacuum.

Substituting the thermal form of $g(v_x)$, one obtains
\begin{equation}
n(x,t)
=
e^{-t/\tau_\mu}
\frac{mx}{kT}
\int_0^t dt_c\,
\frac{F_0(t_c)}{(t-t_c)^2}
\exp\!\left[
-\frac{m x^2}{2kT(t-t_c)^2}
\right].
\end{equation}
The probability that a particle is found in the interval $x_1<x<x_2$ at time $t$ is
\begin{equation}
P(x_1<x<x_2\mid t)
=
\int_{x_1}^{x_2} n(x,t)\,dx.
\end{equation}

\begin{figure}[hbt]
  \centering
  \includegraphics[width=0.9\linewidth]{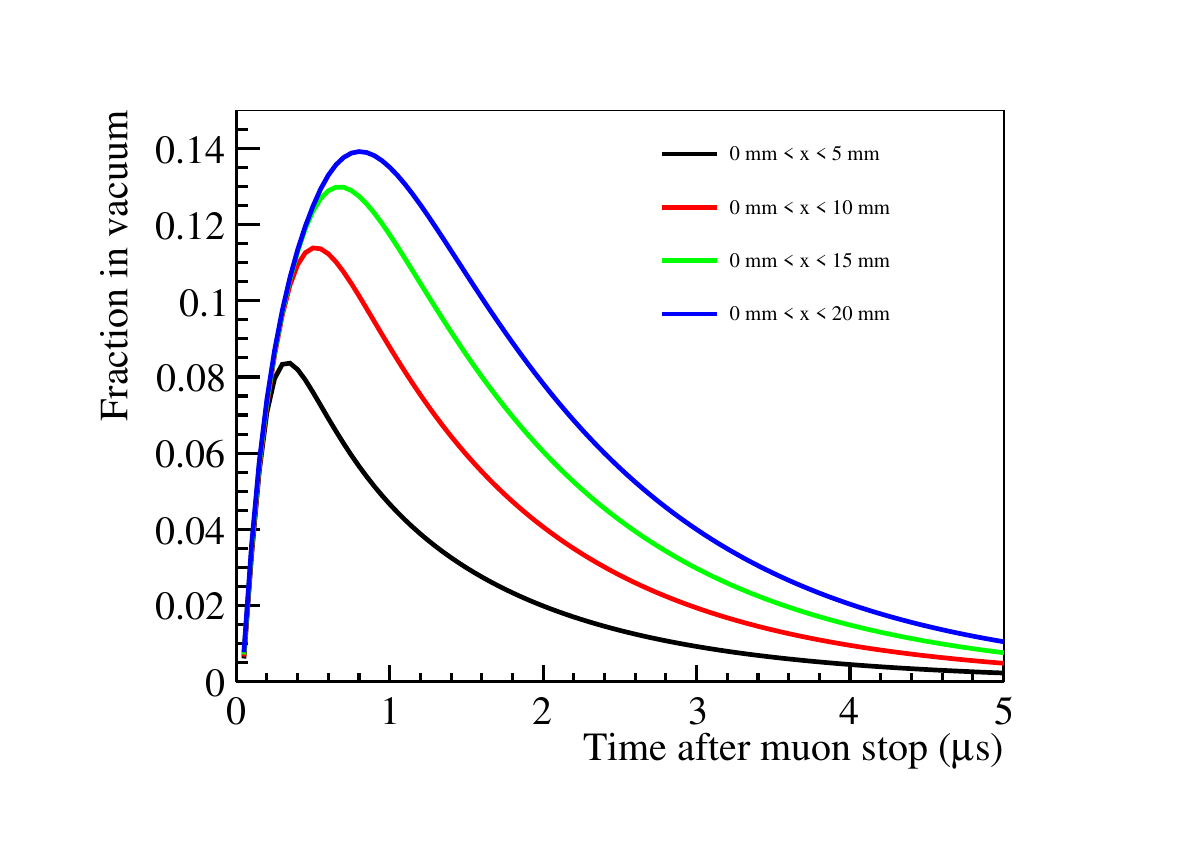}
  \caption{Muonium emission fraction as a function of time.
The vertical axis shows the fraction of muonium that is emitted into vacuum and remains within a given distance from the target surface, normalized to the number of muons stopped within a depth of $L_c$. The horizontal axis shows the time measured from the stopping of the muon in the target. }
  \label{fig:emission-spectrum}
\end{figure}

In practice, the initial stopping distribution is much broader than the characteristic emission length scale, but approximately uniform over the relevant range \(L_c=3\sqrt{D\tau_\mu}\). 
Then the quantity
\begin{equation}
\label{eq:emission}
\int_{x_1}^{x_2} n_c(x,t)\,dx
\end{equation}
directly gives the fraction of particles, relative to those initially stopped within \(L_c\), that are found in the interval \(x_1<x<x_2\) at time \(t\). Here, $n_c(x,t)$ denotes $n(x,t)$ calculated with $L=L_c$.
Figure~\ref{fig:emission-spectrum} shows the time dependence of the fraction of muonium that is emitted into vacuum and remains within a given distance from the target surface, calculated using Eq.~\ref{eq:emission}. 
We assume a hot-tungsten temperature of $T=2000$~K.
The diffusion coefficient is $2\times10^{-3}\,\mathrm{cm^2/s}$~\cite{PhysRevLett.56.1463}. 
For simplicity, the muonium formation efficiency is taken to be unity.

In the multilayer-target configuration, the gap between adjacent layers defines the available free space into which muonium can be emitted. A larger gap is favorable for keeping more muonium within the interlayer region, but it also reduces the number of layers that can be accommodated within a fixed overall system size.

\begin{figure}[t]
  \centering
  \includegraphics[width=0.9\linewidth]{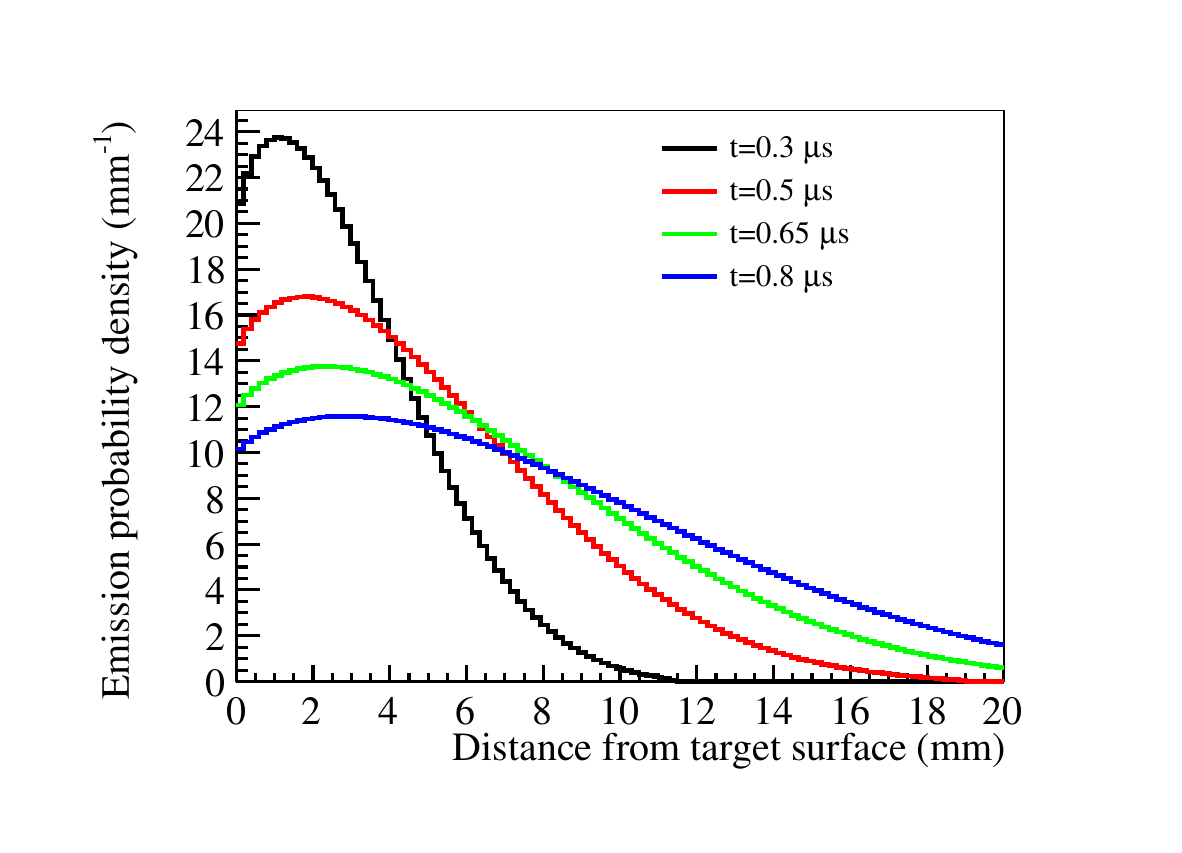}
  \caption{Probability density of emitted muonium at different times after the muon stops in the target. The spatial distribution of ultraslow muons after ionization is expected to reflect these distributions.}
  \label{fig:emission-dist}
\end{figure}

Using the same parameters, Fig.~\ref{fig:emission-dist} shows the probability density of emitted muonium, calculated using Eq.~\ref{eq:emission}, as a function of distance from the target surface for several times after the muon stops in the target. 
The four curves correspond to $t=0.3~\mu\rm s$, $t=0.5~\mu\rm s$, $t=0.65~\mu\rm s$, $t=0.8~\mu\rm s$. These times correspond approximately to those at which the fraction of muonium present in vacuum reaches its maximum for the representative interlayer gaps discussed in Fig.~\ref{fig:emission-spectrum}.

\section{Hadronic-physics model and angle-dependent scale-factor correction}
\label{app:model-comparison}

In this appendix, we describe the choice of the hadronic-physics model in PHITS and the angle-dependent scale-factor correction applied to it.
PHITS provides several physics models for proton-nucleus collisions in the few-GeV energy range, including INCL~\cite{Boudard2013} and JAM~\cite{Nara2000}.
We adopt the JAM model in the present study because its predicted $\pi^+$ momentum distribution shows better agreement with experimental data in the low-momentum region most relevant to the hybrid target.

\begin{figure*}[t]
  \centering
  \begin{minipage}[t]{0.495\linewidth}
    \centering
    \includegraphics[width=0.95\linewidth]{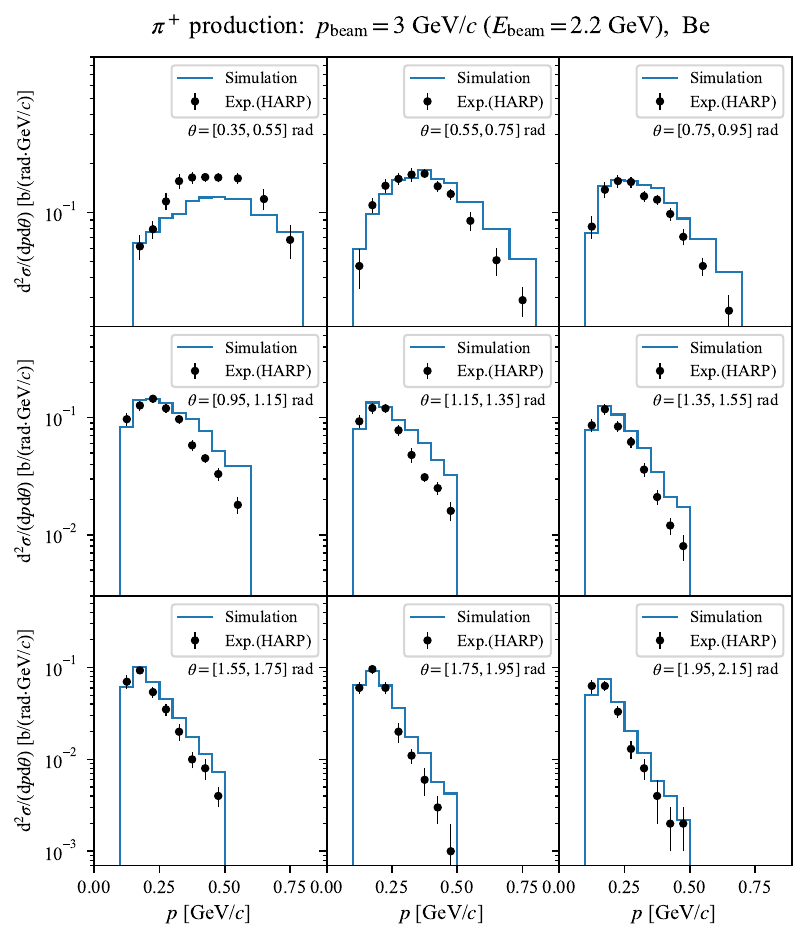}
    \centerline{(a)}
  \end{minipage}%
  \hspace{0.005\linewidth}%
  \begin{minipage}[t]{0.495\linewidth}
    \centering
    \includegraphics[width=0.95\linewidth]{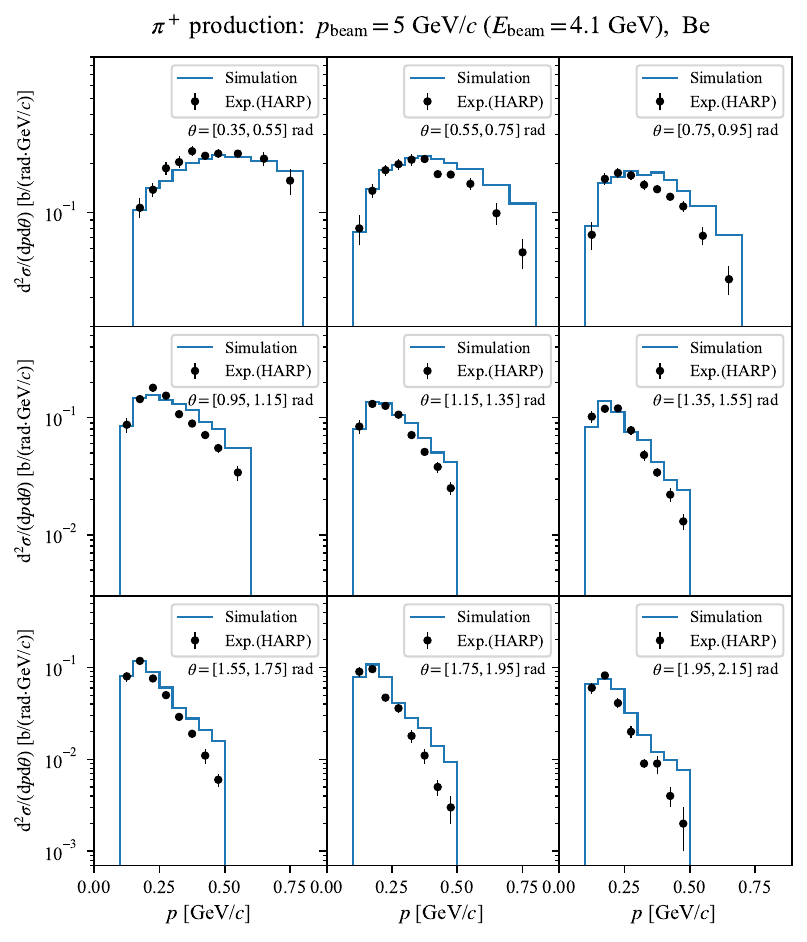}
    \centerline{(b)}
  \end{minipage}
  \caption{Comparison of the double-differential pion-production cross section for a Be target obtained with the JAM model after applying the angle-dependent scale correction and HARP measurements~\cite{HARP2008}. The proton-beam momenta $p_{\rm beam}$ and the corresponding kinetic energies $E_{\rm beam}$ are (a) $3$~GeV/$c$ ($2.2$~GeV) and (b) $5$~GeV/$c$ ($4.1$~GeV).}
  \label{fig:compare-energy}
\end{figure*}

The HARP Collaboration~\cite{HARP2008} measured double-differential pion-production cross sections for various nuclear targets using proton beams with momenta of $p_{\rm beam}=3$--$12.9$~GeV/$c$.
We compare the PHITS simulation results based on the JAM model with the HARP measurements for a Be target at proton-beam kinetic energies of $E_{\rm beam}=2.2$~GeV and $E_{\rm beam}=4.1$~GeV.
These energies bracket the baseline proton-beam energy of 3~GeV assumed in the present study.
The pion-production target in the present study is lithium, whereas no directly corresponding HARP measurement for a Li target is available.
We therefore use the Be data as the closest available light-nucleus benchmark.
Since Li and Be are both low-$Z$ nuclei with similar mass numbers, the residual target-material dependence of the correction is expected to be smaller than the model-normalization discrepancy addressed here.
We nevertheless regard this assumption as part of the hadronic-model systematic uncertainty.
Although the JAM model generally reproduces the measured spectra, a residual normalization discrepancy remains.
We therefore apply an angle-dependent scale factor to the pion-production cross section predicted by the JAM model:
\begin{equation}
f(\theta_{\pi^+}) \;=\; 1.54 - 0.33\,\theta_{\pi^+}/{\rm rad}.
\end{equation}
This correction factor is determined by a fit to the HARP measurements over all angle bins, so that the simulation agrees with the data in the low-momentum region below 200~MeV/$c$.
Figure~\ref{fig:compare-energy} shows the comparison between the prediction by the JAM model after the correction and the result from the HARP measurements.

For pion momenta below 0.3~GeV/$c$, the corrected JAM simulation reproduces the HARP measurements well.
The fit covers the HARP angle bins of $\theta_{\pi^+} = 0.35$--$2.15$~rad and pion momenta below 200~MeV/$c$. Outside these ranges the correction is applied as a linear extrapolation in $\theta_{\pi^+}$, but most of the muonium-producing pions fall within the fit region. The corrected simulation reproduces the HARP data with bin-by-bin residuals of up to $\sim 25\%$ in magnitude. The residuals partially cancel when integrated over angle, leaving the angle-integrated yield within $\sim 10\%$ of the data. We take this as a representative systematic uncertainty of the present hadronic-model treatment.
Residual discrepancies remain in the higher-momentum region above 0.4~GeV/$c$. Their impact on the present study is limited, however, because these high-momentum pions have long ranges and make only a small contribution to the pion-stopping distribution inside the hybrid target.


\bibliography{main}

\end{document}